\documentclass[12pt]{elsarticle}
\makeatletter
\def\ps@pprintTitle{%
	\let\@oddhead\@empty
	\let\@evenhead\@empty
	\def\@oddfoot{\centerline{\thepage}}%
	\let\@evenfoot\@oddfoot}
\makeatother
\usepackage{amsmath,amssymb,hyperref}
\usepackage{amssymb,amsmath,booktabs,hyperref}
\usepackage{graphics}
\usepackage{graphicx}
\usepackage{subfig}
\usepackage{multirow}
\usepackage{tabularx}
\usepackage{rotating}
\usepackage{lscape}
\usepackage{anysize}
\usepackage{xcolor}
\usepackage{listings}
\usepackage{float}
\usepackage{soul}
\usepackage{adjustbox}
\usepackage{colortbl}
\newtheorem{definition}{Definition}[section] 
\usepackage{blindtext,nameref}

\allowdisplaybreaks
\marginsize{2.5cm}{2.5cm}{2.5cm}{2.5cm}
\date{}
\begin{document}
	
	\begin{frontmatter}
		\title{\textcolor{red}{Integrating Fuzzy Set Theory with Pandora Temporal Fault Trees for Dynamic Failure Analysis of Complex Systems}}		
		\author{Hitesh Khungla, Mohit Kumar}

		\address{Department of Basic Sciences,\\ Institute of Infrastructure, Technology, Research And Management, Ahmedabad, Gujarat-380026, India\\
			Email: hitesh.khungla.iitram@gmail.com, mohitkumar@iitram.ac.in}
		
\begin{abstract}
Pandora temporal fault tree, as one notable extension of the fault tree, introduces temporal gates and temporal laws. Pandora Temporal Fault Tree(TFT) enhances the capability of fault trees and enables the modeling of system failure behavior that depends on sequences. The calculation of system failure probability in Pandora TFT relies on precise probabilistic information on component failures. However, obtaining such precise failure data can often be challenging. The data may be uncertain as historical records are used to derive failure data for system components. To mitigate this uncertainty, in this study, we proposed a method that integrates fuzzy set theory with Pandora TFT. This integration aims to enable dynamic analysis of complex systems, even in cases where quantitative failure data of components is unreliable or imprecise. The proposed work introduces the development of Fuzzy AND, Fuzzy OR, Fuzzy PAND, and Fuzzy POR logic gates for Pandora TFT. We also introduce a fuzzy importance measure for criticality analysis of basic events. All events in our analysis are assumed to have exponentially distributed failures, with their failure rates represented as triangular fuzzy numbers. We illustrate the proposed method through a case study of the Aircraft Fuel Distribution System (AFDS), highlighting its practical application and effectiveness in analyzing complex systems. The results are compared with existing results from Petri net and Bayesian network techniques to validate the findings.
\end{abstract}

\end{frontmatter}

\section{Introduction}
Safety-critical systems find extensive application across various industries such as aerospace, automotive, and energy sectors. Failures within these systems possess the capacity to result in significant consequences for both human life and the environment. Thus, a rigorous analysis of system behavior is essential to ensure that these systems maintain a high level of reliability and functionality throughout their operational lifespan. Various techniques have been developed to analyze safety and evaluate system reliability, providing a comprehensive framework for identifying and addressing potential risks. Fault Tree Analysis (FTA)\cite{watson1961launch} is a widely used and well-established technique to compute the reliability of complex systems. In FTA\cite{ftabook}, AND and OR gates are employed to graphically represent the logical relationships between system failure (top event) and its underlying root causes (basic events). This approach enables both qualitative and quantitative analyses, providing a comprehensive understanding of potential system failures. Qualitative analysis is a deductive process that commences with an identified system failure and iteratively works backward to uncover the underlying root causes. Subsequently, boolean logic is applied to minimize the fault tree and to obtain Minimal Cut Sets (MCS) \cite{Fussell1972NEWMF}, which represent the smallest combinations of component failures that can lead to system failure. Once MCSs are identified, quantitative analysis utilizes probabilistic data of the system's components to compute the reliability of the system over a specified time. Quantitative analysis involves computing the probability of each MCS and summing them to obtain the TE probability\cite{Xing2008}.

While FTA is a popular system analysis method, it does have recognized limitations. A notable limitation of FTA is its restricted capability to assess the reliability of static systems only. Static systems are defined by a singular, unchanging mode of operation, whereas modern large-scale systems typically operate in multiple modes or phases, making them dynamic and adaptable. Modern complex systems possess a range of dynamic failure characteristics, including functional dependencies between events, prioritization of failure events and many more. To address this limitation, several extensions to static fault trees have been proposed, including Dynamic Fault Trees (DFTs) \cite{Dugan1992}, Boolean logic Driven Markov Processes (BDMP) \cite{BOUISSOU2003149}, and Pandora Temporal Fault Tree (TFT) \cite{walker2009pandora}. In 1976, Fussell et al.\cite{Fussell} introduced the Priority-AND (PAND) Gate, igniting numerous studies and works focused on exploring its functionalities and applications. The PAND gate stands out among the standard gates due to its ability to impose an order on a set of events, enabling a fault tree to capture time-dependent structures. This feature allows analysts to consider the timing and order of events, enhancing the accuracy and completeness of FTA\cite{WALKER2006237}. Martin Walker and Yiannis Papadopoulos \cite{WALKER200725} introduced the Priority-OR (POR) Gate, which has since been the subject of several developments in the field of temporal gates. Following this development, various case studies have been conducted using TFT to explore its practical applications and effectiveness \cite{ZHOU2022108553} \cite{KABIR2023} \cite{KABIR201620}. 

Fuzzy set theory \cite{ZADEH1965338} has demonstrated effectiveness in solving problems when precise data is unavailable or inaccurate and in making decisions based on vague information \cite{ZADEH1965338} \cite{MISRA1990139} \cite{SURESH1996135}. Tanaka et al. \cite{tanaka1983fault} were the first to apply fuzzy set theory in FTA, modeling BE failure probabilities with trapezoidal fuzzy numbers and employing the fuzzy extension principle to estimate the TE probability. Subsequent extensive research on Fuzzy Fault Tree Analysis (FFTA) was carried out by Misra and Weber \cite{MISRA1990139} \& Liang and Wang \cite{LIANG1993583}. In addition, fuzzy set theory combined with expert elicitation was utilized by Ching-Torng Lin and Mao-Jiun J. Wang \cite{LIN1997205} to assess the reliability of a robot drilling system. The application of FFTA has been used to assess the reliability of different systems. For example, Yuhua and Datao \cite{YUHUA200583} employed FFTA to assess the likelihood of failure in an oil and gas transmission system. Deng-Feng Li \cite{LI20081741} utilized FFTA based on intuitionistic fuzzy sets to analyze failures in printed circuit board assembly. Ferdous et al. \cite{Ferdous} proposed a computer-aided fuzzy fault tree analysis method. S. Rajakarunakaran et al.\cite{RAJAKARUNAKARAN2015} apply FFTA and expert elicitation method to evaluate risk in LPG refueling station. Hailong Yin et al.\cite{YIN2020} evaluated the safety of natural gas storage tanks using a method called Similarity Aggregation Method based Fuzzy Fault Tree Analysis (SAM-FFTA). Singh et al.\cite{SINGH2022189} developed $\alpha$-cut interval-based FFTA with Bayesian network for criticality analysis of submarine pipeline leakage. K. Singh et al.\cite{SINGH2022189} developed $\alpha$-cut interval-based FFTA with Bayesian network for criticality analysis of submarine pipeline leakage. Kaushik M. and Kumar M. \cite{kaushik2023integrated} proposed an integrated approach of intuitionistic fuzzy fault tree and Bayesian network analysis applicable to risk analysis of ship mooring operations.

While a significant amount of research has focused on using fuzzy set theory in classical FTA to enable quantitative analysis, there has been limited exploration in extending this capability to TFTA. Recently, initial concepts regarding fuzzy set theory based Pandora TFTA were introduced by Kabir et. al.\cite{kabir2014}. The existing article revealed gaps in the application of conventional formulas to temporal gates like PAND and POR. To overcome these challenges, we derived new formulas for PAND and POR gates and employed them to compute the TE probability. In this paper, we have developed an algorithm to compute the TE probability of TFTA using fuzzy failure rates. Here, we have taken failure rates as triangular fuzzy numbers. we have compared our results with other existing methods such as Petri net \cite{CODETTARAITERI200545} and Bayesian network\cite{MONTANI2008922} \cite{BOBBIO2001249}. We applied our algorithm to an AFDS and ranked each BE according to its impact on the probability of TE using fuzzy importance measures.

The paper is structured as follows: Section 2 covers the basic definitions of fuzzy set theory and formulas for pandora Gates in classical set theory, and Section 3 extends those formulas in a fuzzy environment and ranking has been conducted using a Euclidean Distance-based Fuzzy Importance Measure (FIM). In Section 4, we present a case study of an AFDS where we compute the TE probability using the proposed algorithm and then we conclude the paper by comparing our results with existing methods such as Petri net-based solutions and Bayesian network-based solutions. 

\section{Background}
This section presents fundamental definitions of fuzzy set theory and provides a basic overview of Pandora TFT.
\begin{definition}
Fuzzy Number \cite{ZADEH1965338}
\end{definition}

A fuzzy set $\tilde{A}$ defined on universal set $\mathbb{X}$, is an ordered pair $\tilde{A} = \left\lbrace <x, \mu_{\tilde{A}}(x)> : x \in \mathbb{X} \right\rbrace $, where $\mu_{\tilde{A}}(x) : \mathbb{X} \rightarrow  [0,1] $ is membership function of set $\tilde{A}$.  \\ \\
A fuzzy set $\tilde{A}$ defined on $\mathbb{R}$ is called fuzzy number if it possess the following conditions
\begin{enumerate}
    \item $\tilde{A}$ is normal, i.e. $\exists x \in \mathbb{R} \text{ s.t. } \mu_{\tilde{A}}(x)=1$
    \item $\tilde{A}$ is convex, i.e. $\forall x_1,x_2 \in \mathbb{R}, 0 \le \lambda \le 1$
    \begin{equation*}
        \mu_{\tilde{A}}(\lambda x_1 + (1-\lambda)x_2) \ge min\{\mu_{\tilde{A}}(x_1),\mu_{\tilde{A}}(x_2)\}
    \end{equation*}
    \item supp$(\tilde{A})=\{ x \in \mathbb{R}: \mu_{\tilde{A}}(x)>0 \}$ is bounded.
\end{enumerate}

A fuzzy number $\tilde{A} = \left\lbrace < x, \mu_{\tilde{A}}(x) > \: x \in \mathbb{R} \right\rbrace $ is said to be a Triangular Fuzzy Number (TFN)\cite{klir} if membership function is defined as follows:

\begin{equation*}
	\mu_{\tilde{A}}(x) = 
	\left\{\begin{matrix}
		\frac{x-a}{b-a} \ \ \ if \ a \leq x \leq b,\\
		\\
            \frac{c-x}{c-b} \ \ \ if \  b \leq x \leq c,\\
		\\
		 \ \ 0\ \ \ \ \ \ \     \text{otherwise}.
	\end{matrix}\right.
\end{equation*}
TFN is denoted by $\tilde{A}=(a,b,c)$.

\begin{definition}
    Arithmetic Operations on Two Triangular Fuzzy Numbers \cite{DIDIER1978}
\end{definition}
Let $\tilde{A}=(a_1,a_2,a_3)$ and $\tilde{B}=(b_1,b_2,b_3)$ be two TFNs. For $a_1,b_1>0$ and for any scalar $k$, arithmetic operations on $\tilde{A}$ and $\tilde{B}$ are defined as follows: 
\begin{enumerate}
    \item  $\tilde{A}+\tilde{B} =(a_1+b_1,a_2+b_2,a_3+b_3)$
    \item $\tilde{A}-\tilde{B} =(a_1-b_1,a_2-b_2,a_3-b_3)$ 
    \item $\tilde{A}*\tilde{B} =(a_1*b_1,a_2*b_2,a_3*b_3)$  
    \item  $\tilde{A}/\tilde{B} =(a_1/b_3,a_2/b_2,a_3/b_1)$ 
    \item  $e^{k\tilde{A}} = \begin{cases} 
        (e^{ka_1},e^{ka_2},e^{ka_3}) & \text{if} \  k>0 \\
        (e^{ka_3},e^{ka_2},e^{ka_1}) & \text{if} \  k<0
                    \end{cases}$ 
\end{enumerate}
   
\begin{definition}
    Defuzzification \cite{1993defuzzification}
\end{definition}
Defuzzification is the process of converting fuzzy set outputs into crisp values. In fuzzy logic systems, the output is often a fuzzy set representing linguistic variables or terms. Defuzzification involves mapping these fuzzy sets to crisp values that can be easily interpreted or used in decision-making processes. There are many  defuzzification methods are available in literature\cite{HELLENDOORN1993} \cite{FORTEMPS1996} \cite{LEEKWIJCK1999} \cite{OPRICOVIC2003}. In this paper, we use the centroid method, defined as follows: \\ \\
For any fuzzy set $\tilde{A}$ with membership function $\mu_{\tilde{A}}(x)$,
\begin{equation*}
    d(\tilde{A})=\frac{\int_{\mathbb{R}}x\mu_{\tilde{A}}(x) dx}{\int_{\mathbb{R}}\mu_{\tilde{A
    }}(x) dx}
\end{equation*}
If $\tilde{A}=(a,b,c)$ is a TFN then defuzzification of $\tilde{A}$ is given by $\frac{a+b+c}{3}$.  

\subsection{Pandora Temporal Fault Tree Analysis}
The Pandora TFT expands upon traditional FTA to overcome its limitations in analyzing dynamic systems. Pandora TFT differs from FTA by integrating temporal aspects such as time-dependent events, event sequences, and dynamic behaviors. This extension introduces temporal gates and laws to accurately represent temporal relationships and dependencies between events. Consequently, Pandora TFT enables the analysis of system failures over time, offering a comprehensive perspective on system behavior beyond static considerations. While FTA utilizes only the AND and OR gates, Pandora TFTs introduce two new gates, namely PAND and POR. It is beneficial for understanding the dynamic behavior of complex systems. Pandora TFTs also employ the Boolean AND and OR gates. The symbols $\cup$ and $\cap$ are utilized to denote the OR and AND gates respectively, in logical expressions. For the PAND and POR gate, $\lhd$ and $\wr$ symbols are used respectively. In the quantitative analysis of Pandora TFTs, the probability of TE can be estimated using the failure rates of BEs as input. Quantifying TFTA involves evaluating the gates quantitatively. 

The AND gate symbolizes a logical AND operation, indicating that the output event occurs only when all the input events occur. Let events $A_1$ to $A_n$ be independent of each other and connected with event A using AND gate. Then the probability of event A is defined as follows:\cite{hoyland}
    \begin{equation*}
   Pr(A)=Pr(A_1 \cap A_2 \cap ... \cap A_n)=\prod_{i=1}^{n}Pr(A_i)
    \end{equation*} 

Similarly, the OR gate represents a logical OR operation. It represents a situation where the output event will occur if at least one of the input events occurs. Let events $A_1$ to $A_n$ be independent of each other and connected with event A using OR gate. Then the probability of event A is defined as follows:\cite{hoyland}
    \begin{equation*}
    Pr(A)=Pr(A_1 \cup A_2 \cup ... \cup A_n)=1-\prod_{i=1}^{n}(1-Pr(A_i))
    \end{equation*}

Also, we have two temporal logic gates PAND and POR. A PAND gate is a logic gate, which is equivalent to an AND gate but with the requirement that the input events must occur in a specific sequence. Event $A$ is the result of the simultaneous occurrence of all events $A_n$, $A_{n-1}$,$...$,$A_1$. Moreover, $A$ only happens if and only if event $A_n$ occurs first, followed by event $A_{n-1}$ second, and so on until event $A_1$ occurs last. Conventionally, the events connected to a PAND gate must unfold in the sequence they are arranged from left to right.

Let $\lambda_i$ be a failure rate for basic event $A_i$ and suppose that all basic events are non-repairable then the probability of PAND gate is defined as follows:\cite{Fussell}
\begin{equation}
     Pr(A_n \lhd A_{n-1} \lhd ... \lhd A_1)=\prod_{i=1}^{n}\lambda_i \sum_{k=0}^{n} \left[ \frac{e^{a_kt}}{\prod_{j=0,j \neq k}^{n}(a_k - a_j)} \right]
\end{equation}
where, 
$a_0=0$ and $a_m=-\sum_{j=1}^{m} \lambda_j$ for $m>0$. \\ 

Similarly, A Priority-OR (POR) gate extends the functionality of an OR gate by introducing additional conditions. The term 'priority' is employed because the order of events is considered, unlike in a simple OR gate where all inputs can trigger the output. Therefore, the expression A POR B is satisfied if either A occurs before B (with the possibility of both occurring), or if A occurs and B does not occur. Let events $A_n,A_{n-1},...,A_1$ are connected with the POR gate then the probability of the POR gate is defined as follows:\cite{EDIFOR2012}
\begin{equation}
    Pr(A_n \wr A_{n-1} \wr ... \wr A_1)=\frac{\lambda_1 \left( 1- e^{-(\sum_{i=1}^{n}\lambda_i)t} \right)}{\sum_{i=1}^{n}\lambda_i}
\end{equation}

\section{Fuzzy Pandora TFTA} 
In this section, we are going to integrate fuzzy set theory with the logic gates of Pandora TFT. We develop fuzzy formulas for Pandora TFTA operators.
 
\subsection{Fuzzy AND Gate and Fuzzy OR Gate}
    Let the events $A_1,A_2,...A_n$ be connected with the AND gate and consider the probability of each event $A_i$ as a TFN $(a_i,b_i,c_i)$ the fuzzy probability for the fuzzy AND gate is defined as follows:
    
\begin{equation}
    Pr(A_1 \cap A_2 \cap ... \cap A_n)=\left( \prod_{i=1}^{n}a_i,\prod_{i=1}^{n}b_i,\prod_{i=1}^{n}c_i \right)
\end{equation}

Similarly, If all these events are connected with the OR logic gate then the fuzzy probability for the fuzzy OR gate is defined as follows:

\begin{equation}
    Pr(A_1 \cup A_2 \cup ... \cup A_n)=\left( 1-\prod_{i=1}^{n}(1-a_i),1-\prod_{i=1}^{n}(1-b_i),1-\prod_{i=1}^{n}(1-c_i) \right)
\end{equation}

\subsection{Fuzzy PAND Gate}
    Let the failure rate of an event $A_i$ be a TFN $(a_i,b_i,c_i)$ then the fuzzy probability of event $A_i$ is defined as follows:
\begin{equation*}
    Pr(A_i)=\int_{0}^{t_i} (a_i,b_i,c_i) e^{-(a_i,b_i,c_i)t_i} dt_i.
\end{equation*}
Therefore,
\begin{equation}
    Pr(A_i)=\left(\int_{0}^{t_i} a_i e^{-c_i t_i} dt_i,\int_{0}^{t_i} b_i e^{-b_i t_i} dt_i,\int_{0}^{t_i} c_i e^{-a_i t_i} dt_i\right)
    \label{praif}
\end{equation}
But,
\begin{equation*}
\begin{split}
    \int_{0}^{t_{i-1}}f_i(t_i) dt_i & = \int_{0}^{t_{i-1}} (a_i e^{-c_i t_i},b_i e^{-b_i t_i},c_i e^{-a_i t_i}) dt_i \\
    & = \left(\int_{0}^{t_{i-1}} a_i e^{-c_i t_i} dt_i,\int_{0}^{t_{i-1}} b_i e^{-b_i t_i} dt_i,\int_{0}^{t_{i-1}} c_i e^{-a_i t_i} dt_i \right) \\
    & = \left( L^{-1}\left( \frac{a_i}{s(s+c_i)} \right), L^{-1}\left( \frac{b_i}{s(s+b_i)} \right), L^{-1}\left( \frac{c_i}{s(s+a_i)} \right) \right)
\end{split}
\end{equation*}
where, $L^{-1}$ is the inverse Laplace transform operator. \\

Let us consider that the events $A_n,A_{n-1},...,A_1$ are connected with the PAND gate from left to right. Denote $A=A_n \lhd A_{n-1} \lhd ... \lhd A_1$ Then,
\begin{equation}
    Pr(A)=\int_{0}^{t}f_1(t_1) \int_{0}^{t_1}f_2(t_2) \int_{0}^{t_2} ... \int_{0}^{t_{n-1}}f_n(t_n)  dt_n ... dt_2 dt_1
    \label{pandeq}
\end{equation}
where
$f_i(t_i)=\lambda_i e^{-\lambda_i t_i} = (a_ie^{-c_it_i},b_ie^{-b_it_i},c_ie^{-a_it_i})$\cite{Fussell}. After using the inverse Laplace transform we have,

\begin{equation*}
Pr(A)=\left( L^{-1} \left\{ \frac{1}{s}\prod_{j=1}^{n} \left( \frac{a_j}{s-p_j} \right) \right\},L^{-1} \left\{ \frac{1}{s}\prod_{j=1}^{n} \left( \frac{b_j}{s-q_j} \right) \right\},L^{-1} \left\{ \frac{1}{s}\prod_{j=1}^{n} \left( \frac{c_j}{s-r_j} \right) \right\}  \right)
\end{equation*}
where, $p_0=q_0=r_0=0$ and for $m>0$,
\begin{equation*}
    p_m=-\sum_{i=1}^{m}c_i, \ \ \ q_m=-\sum_{i=1}^{m}b_i, \ \ \ r_m=-\sum_{i=1}^{m}a_i.
\end{equation*}
Using Heaviside's expansion formula,
\begin{multline*}
    Pr(A)=\bigg( \prod_{i=1}^{n}a_i \sum_{k=0}^{n} \left[ \frac{e^{p_kt}}{\prod_{j=0.j \neq k}^{n}(p_k - p_j)} \right],\prod_{i=1}^{n}b_i \sum_{k=0}^{n} \left[ \frac{e^{q_kt}}{\prod_{j=0.j \neq k}^{n}(q_k - q_j)} \right], \\ \prod_{i=1}^{n}a_i \sum_{k=0}^{n} \left[ \frac{e^{r_kt}}{\prod_{j=0.j \neq k}^{n}(r_k - r_j)} \right] \bigg).
\end{multline*}
For $n=2$, the above equation can be rewritten as follows, 
\begin{multline*}
    Pr(A_1 \lhd A_2) = \bigg( a_1a_2 \left[ \frac{1}{c_1(c_1+c_2)} - \frac{e^{-c_1t}}{c_1c_2} + \frac{e^{-(c_1+c_2)t}}{c_2(c_1+c_2)} \right],b_1b_2 \bigg[ \frac{1}{b_1(b_1+b_2)} - \frac{e^{-b_1t}}{b_1b_2} \\ + \frac{e^{-(b_1+b_2)t}}{b_2(b_1+b_2)} \bigg],  c_1c_2 \left[ \frac{1}{a_1(a_1+a_2)} - \frac{e^{-a_1t}}{a_1a_2} + \frac{e^{-(a_1+a_2)t}}{a_2(a_1+a_2)} \right] \bigg).
\end{multline*}

\subsection{Fuzzy POR Gate}
If events $A_n,A_{n-1},...,A_1$ are connected with the POR gate from left to right. Denote $A=A_n \wr A_{n-1} \wr ... \wr A_1 $ then,

\begin{equation}
Pr(A)=\int_{0}^{t}\left( f_{A_n}(x) \left( 1-F_{A_{n-1}}(x) \right) ...\left(1-F_{A_1}(x) \right) \right) dx
\label{5}
\end{equation} 
where, 
$ F_{A_i}(x) = \int_{0}^{t_i} f_{A_i}(x) dx $ and $f_{A_i}(x)=\lambda_i e^{-\lambda_i x}$. \\ 
For a fuzzy failure rate $\lambda_i$ we have,
\begin{equation}
\begin{split}
    F_{A_i}(t) & = \int_{0}^{t}\lambda_i e^{-\lambda_i x} dx \\
            & = \int_{0}^{t} \left( a_ie^{-c_ix},b_ie^{-b_ix},c_ie^{-a_ix}  \right) dx \\
            & = \left( \frac{a_i}{c_i}-\frac{a_i}{c_i}e^{-c_it}, 1-e^{-b_it}, \frac{c_i}{a_i}-\frac{c_i}{a_i}e^{-a_it} \right).
\end{split}
\label{6}
\end{equation}
From eq.\ref{5} and eq.\ref{6},
\begin{multline*}
    Pr(A)=\bigg( \int_{0}^{t} (a_ne^{-c_nx})(1-\frac{c_{n-1}}{a_{n-1}}+\frac{c_{n-1}}{a_{n-1}}e^{-a_{n-1}x})...(1-\frac{c_1}{a_1}-\frac{c_1}{a_1}e^{-a_1x}) dx,\\ \int_{0}^{t} b_n (1-e^{-(b_n+b_{n-1}+...+b_1)x}) dx, \int_{0}^{t} (c_ne^{-a_nx})(1-\frac{a_{n-1}}{c_{n-1}}+\frac{a_{n-1}}{c_{n-1}}e^{-c_{n-1}x})...(1-\frac{a_1}{c_1}-\frac{a_1}{c_1}e^{-c_1x}) dx \bigg).
\end{multline*}
For n=2, the above equation can be rewritten as follows,
\begin{multline*}
  Pr(E_1 \wr E_2)=  \bigg( \frac{a_1c_2}{a_2(c_1+a_2)}(1-e^{-(c_1+a_2)t}) + \frac{a_1(a_2-c_2)}{a_2c_1}(1-e^{-c_1t}) , \frac{b_1}{b_1+b_2}(1-e^{-(b_1+b_2)t}), \\ 
  \frac{c_1a_2}{c_2(a_1+c_2)}(1-e^{-(a_1+c_2)t}) + \frac{c_1(c_2-a_2)}{c_2a_1}(1-e^{-a_1t}) \bigg).  
\end{multline*}

\subsection{Proposed Algorithm}
In this section, we utilize fuzzy formulas from Pandora TFTA to calculate the TE probability. Additionally, we discuss the importance measures for each BE, providing insights into their relative contributions to system reliability.

\subsubsection{Obtaining Failure Data of Components}
Determining the failure rate of components involves collecting information regarding the frequency of failures occurring within a given period of time. We can gather failure rate data from various sources like field observations, historical records and from databases containing information about reliability.

\begin{itemize}
    \item Field observations involve directly monitoring the performance of components in real-world operating conditions. During routine inspections or maintenance activities, field engineers or technicians may record instances of failures as they occur.
    \item Historical records of component failures provide valuable data within an organization or industry. They highlight trends, patterns, and common failure modes, guiding future failure rate estimates.
    \item Reliability databases compile failure data from various sources such as field observations, testing results, manufacturer specifications and historical records. They provide a platform through which we can observe the failure rate of components.
\end{itemize}

\subsubsection{Fuzzification of Failure Rate}
Fuzzification is the process of converting a precise, crisp value into a fuzzy value, which is represented by a range of possibilities or a degree of membership in a set, rather than a single exact value. Let $\lambda$ represent any crisp failure rate, we can fuzzify it by applying a spread of $\pm 15 \%, \pm 25 \%,$ or $ \pm 50 \%.$ For example, if we choose a $ \pm 15 \%$ spread, the fuzzy failure rate $\tilde{\lambda}$ is defined as follows:\cite{NAVEEN2017}
\begin{equation*}
    \tilde{\lambda}=(0.85*\lambda,\lambda,1.15*\lambda)
\end{equation*}

\subsubsection{Calculating Top Event Probability} 
To determine the TE probability, we initially select a specific time $t$. Subsequently, we calculate the fuzzy failure probability of each component based on its fuzzy failure rate. Finally, employing formulas of different gates, we compute the TE probability. Sometimes, to calculate the probability of the TE, it is necessary to convert failure rates into failure probabilities and failure probabilities into failure rates, which can be achieved using the following formulas.

\begin{itemize}
    \item Convert Failure Rate into Failure Probability: \\
    If event $E$ has exponential failure distribution with failure rate $\lambda$, then probability density function of E is given by $f(t)=\lambda e^{-\lambda t}$ then, 
\begin{equation}
    Pr(E)=\int_{0}^{t}\lambda e^{-\lambda t} dt = 1-e^{-\lambda t}.
    \label{eq1}
\end{equation}

    \item Convert Failure Probability into Failure Rate: \\
    Let $E$ be the event with probability $P$, then from eq.\ref{eq1}, Failure rate $\lambda$ of event $E$ is given by, \cite{KABIR201620}
\begin{equation}
    \lambda = \frac{-ln(1-P)}{t}
    \label{eq2}
\end{equation}
\end{itemize}

\subsubsection{Importance Analysis} 
Importance measures evaluate the different impacts of basic or intermediate events on the likelihood of the TE occurring, or how changes in these events can affect the occurrence of the TE. This data serves as valuable information for resource allocation decisions (such as upgrades, maintenance, etc.) and assists stakeholders in enhancing system dependability, including safety, reliability, and availability. In classical FTA, the Fussell-Vesely and Birnbaum importance measures \cite{meng2000relationships} are widely employed in quantitative evaluations, which assume constant failure rates of components. In this section, a similar importance measure technique is presented, which applies to systems with components having fuzzy failure data.

We determine the fuzzy importance of a BE by computing the difference between the fuzzy probabilities of the TE with and without the presence of the BE. Let $P_{T_{i=1}}$ be the fuzzy probability of TE with probability of BE $E_i$ is $(1,1,1)$ and $P_{T_{i=0}}$ be the fuzzy probability of TE with probability of BE $E_i$ is $(0,0,0)$. The Fuzzy Importance Measure (FIM) is defined as a Euclidean Distance (ED) between above two fuzzy numbers\cite{KABIR201620}, i.e.

\begin{equation*}
    FIM(E_i)=ED[P_{T_{i=1}},P_{T_{i=0}}].
\end{equation*}

Utilizing the above equation, we can compute the importance measure for all the BEs and arrange them based on their importance index.
For two BE $E_i$ and $E_j$, if we have $FIM(E_i)>FIM(E_j)$, 
In such a scenario, the $E_i$ would hold more significance compared to $E_j$.

\section{Case study}
This section illustrates the application of our proposed algorithm to the Aircraft Fuel Distribution System, calculating TE probability at different time values and assessing BE criticality through importance measures.

\subsection{System Description}
The system comprises 2 engines, 5 fuel tanks for fuel storage, 7 bi-directional fuel pumps equipped with speed sensors for fuel distribution, 11 valves to regulate pathways as per system requirements, 6 flow meters to measure fuel flow rates, a refueling point, and 2 jettison points for fuel release if necessary. The compositional analysis is facilitated by dividing AFDS into three subsystems: Port Feed (PF), Central Reservation (CR), and Starboard Feed (SF), with PF and SF featuring identical component arrangements. SF is organized into Starboard Inner Subsystem (SIS) and Starboard Outer Subsystem (SOS), with SIS specifically encompassing the Starboard Inner Tank (SIT), fuel level sensor, and valve SIV.

The AFDS primarily performs two operations, one is fuel storage and another one is the distribution of fuels from tanks. \\
1. Refueling Phase: Fuel injection into the Central Reservation Tank (CRT), followed by automated distribution to the Port Outer Tank (POT), Starboard Outer Tank (SOT), and other tanks. \\
2. Consumption Phase (taxiing, take-off, cruising, approaching, landing): Engines consume fuel; the Port Outer Subsystem (POS) supplies the Port Engine (PE) from POT, while the Starboard Outer Subsystem (SOS) feeds the Starboard Engine (SE) from SOT. Within SF, valve SIV in SIS remains closed.

\begin{figure}[H]
    \centering
    \includegraphics[scale=0.50]{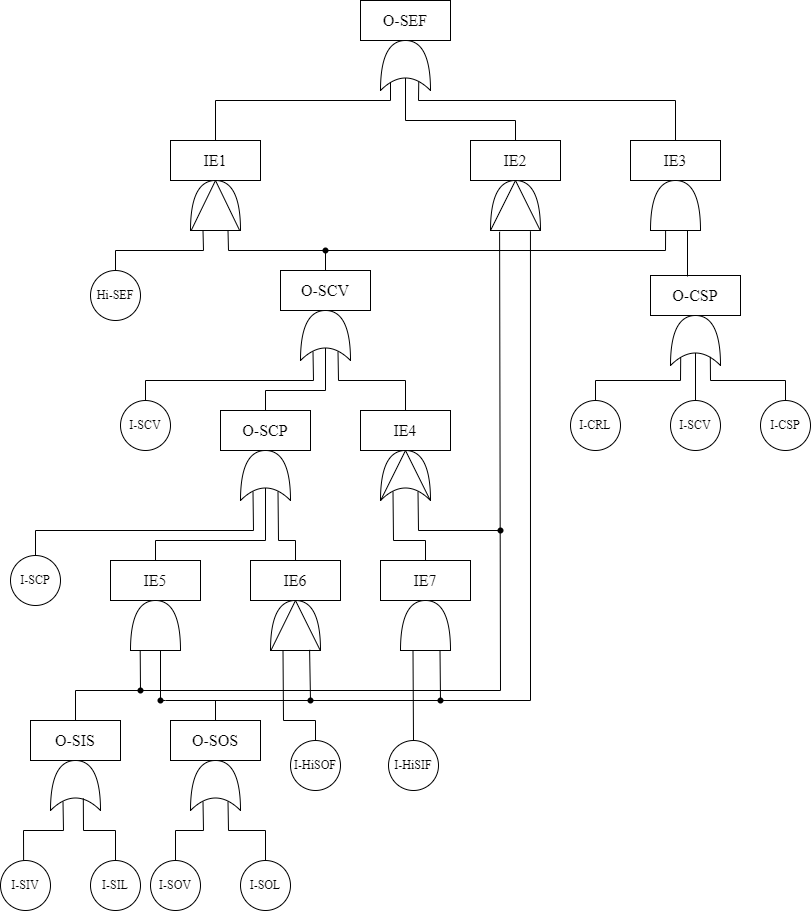}
    \caption{Pandora TFT of the AFDS}
    \label{fta}
\end{figure}

If fuel flow from the SOS is disrupted for any reason, the SIS is activated (while the SOS is deactivated), and fuel is drawn from the SIT to supply the SE. If the SIS fails to provide fuel to the SE, the entire SF subsystem is deactivated, and fuel is drawn from the CRT in the CR subsystem. Similar operations are possible within the PF subsystem. Figure \ref{fta} illustrates a Pandora TFT of AFDS. In Figure \ref{fta}, O-SEF represents a flow meter managing fuel flow at SE.

\subsection{Proposed Algorithm}
In this section, we demonstrate the effectiveness of our proposed algorithm through a case study on the AFDS. We apply the algorithm to compute TE probability at various time intervals and evaluate the criticality of each BE, ranking them according to their importance.

\subsubsection{Obtaining Failure Rate of Each BEs}
Based on various observations, expert opinions, and historical records, Edifor et al.\cite{EDIFOR2014} derived the failure rate $\lambda$ (per hour) of each BE of AFDS, as presented in Table \ref{t0}.

\begin{table}[H]
    \centering
    \begin{tabular}{c|c|c} \hline
        Basic Events & Descriptions & Failure Rate  \\ \hline
        I-SCP & Pump between SOT and SE & 5.84267E-5 \\
        I-CSP & Pump between CRL and SE & 5.84267E-5 \\
        I-SOV & Valve inside SOS & 1.65633E-3 \\
        I-SIV & Valve inside SIS & 1.65633E-3 \\
        I-CSV & Valve between CRL and SE & 1.65633E-3 \\
        I-SCV & Valve between SOT and SE & 1.65633E-3 \\
        I-CRL & Level sensor of CRT & 2.21127E-6 \\
        I-HiSOF & Flow meter between SOT and SE & 4.06861E-5 \\
        I-HiSIF & Flow meter between SIS and SE & 4.06861E-5 \\
        I-HiSEF & Flow meter between SCV and SE & 4.06861E-5 \\
        I-SIL & Level sensor of SIT & 1.65633E-3 \\
        I-SOL & Level sensor of  SOT & 3.31774E-5 \\ \hline
    \end{tabular}
    \caption{Failure Rate of Basic Events}
    \label{t0}
\end{table}

\subsubsection{Fuzzification of Failure Rate of Each BEs}
Due to uncertainties arising from various factors, the crisp failure rates need to be fuzzified to mitigate uncertainty. 
We have utilized a $\pm15\%$ spread to fuzzify the failure rate.

\begin{table}[H]
    \centering
    \begin{tabular}{c|c} \hline
        Basic Events  & Fuzzy Failure Rate  \\ \hline
        I-SCP & (4.96627E-5,5.84267E-5,6.71907E-5) \\
        I-CSP & (4.96627E-5,5.84267E-5,6.71907E-5) \\
        I-SOV & (1.40788E-3,1.65633E-3,1.90478E-3) \\
        I-SIV  & (1.40788E-3,1.65633E-3,1.90478E-3) \\
        I-CSV  & (1.40788E-3,1.65633E-3,1.90478E-3) \\
        I-SCV & (1.40788E-3,1.65633E-3,1.90478E-3) \\
        I-CRL & (1.87958E-6,2.21127E-6,2.54296E-6) \\
        I-HiSOF  & (3.45832E-5,4.06861E-5,4.67889E-5) \\
        I-HiSIF & (3.45832E-5,4.06861E-5,4.67889E-5) \\
        I-HiSEF & (3.45832E-5,4.06861E-5,4.67889E-5) \\
        I-SIL  & (1.40788E-3,1.65633E-3,1.90478E-3) \\
        I-SOL & (2.82008E-5,3.31774E-5,3.81541E-5) \\ \hline
    \end{tabular}
    \caption{Fuzzy Failure Rate of Basic Events}
    \label{t1}
\end{table}

\subsubsection{Evaluating TE Probability}
To compute the probability of TE, we require the failure rate of each BE. Table \ref{t0} and Table \ref{t1} provides descriptions of BEs and their corresponding fuzzy failure rates. From Figure \ref{fta}, the failure probability of O-SEF needs to be computed. To achieve this, we utilize the failure rate of BEs, converting it into failure probability using equation \ref{eq1}. For O-SOS, the calculation is as follows:
\begin{equation*}
    Pr(O-SOS)=Pr(I-SOV \cup I-SOL).
\end{equation*}
After calculating the Pr(O-SOS), we can convert it into a failure rate using equation \ref{eq2}. For IE6, the calculation of probability is as follows:
\begin{equation*}
    Pr(IE6)=Pr(I-HiSOF \lhd O-SOS).
\end{equation*}
Similarly, Pr(O-SEF) can be computed as follows:
\begin{equation*}
    Pr(O-SEF)=Pr(IE1 \cup IE2 \cup IE3).
\end{equation*}

 From the above equation, we have calculated the probability of O-SEF for different values of time $t$, which is given in Table \ref{tep} and we can observe that as time $t$ increases, the failure probability of O-SEF also increases. 

\begin{table}[H]
    \centering 
    \begin{tabular}{c|c} \hline
        t (hours) & Pr(TE) \\ \hline
        100  & 0.05269844 \\
        500  & 0.56162025 \\
        1000 & 0.85253225 \\
        1500 & 0.94286331 \\
        2000 & 0.97599884 \\
        2500 & 0.98951921 \\
        3000 & 0.99532702 \\ 
        \hline
    \end{tabular}
    \caption{Probability of TE for different values of t}
    \label{tep}
\end{table}

\subsubsection{Quantifying Importance Measure of Each BEs}
In addition to analyzing the probability of the TE, ranking each BE based on its importance to the TE is also crucial. In this study, we have calculated the FIM measure for each BE given in Table \ref{fimbe} and ranked them accordingly based on their importance.
From Table \ref{fimbe}, we observe that I-CSV is the most critical BE, while I-HiSOF is the least critical event. Additionally, I-SIV and I-SIL have the same importance measure, placing them at equal rank.
\begin{table}[H]
    \centering
    \begin{tabular}{c|c|c} \hline
         Basic Event & FIM(BE) & Rank  \\ \hline
        I-CSV & 1.2744 & 1 \\
        I-SOV & 0.9774 & 2 \\
        I-SOL & 0.6008 & 3 \\
        I-CSP & 0.5723 & 4 \\
        I-CRL & 0.5566 & 5 \\
        I-SCV & 0.4131 & 6 \\
        I-SCP & 0.1943 & 7 \\ 
        I-SIV & 0.0734 & 8 \\
        I-SIL & 0.0734 & 8 \\
        I-HiSIF & 0.0535 & 9 \\
        I-HiSEF & 0.0022 & 10 \\
        I-HiSOF &  0.0003 & 11 \\ \hline
    \end{tabular} 
    \caption{Fuzzy Importance Measure of BEs}
    \label{fimbe}
\end{table}

\subsection{Result Comparison with Existing Methods}
\begin{figure}[H]
    \centering
    \includegraphics[scale=0.8]{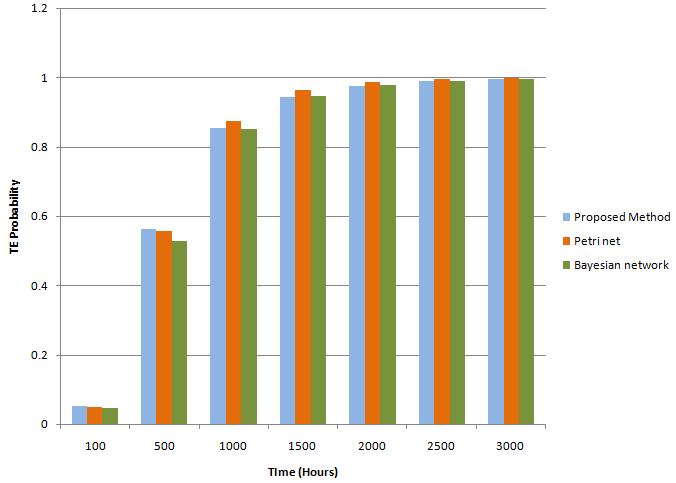}
    \caption{Comparison of TE probability estimated by PN, BN and proposed method}
    \label{g1}
\end{figure}

In this section, we have compared our derived probability of the TE with other existing methods, such as Petri-net based solutions and Bayesian network based solutions. Kabir et. al.\cite{KABIR201855} conducted a study on the same case using crisp data and calculated the probability of the TE using PN and BN. From Table \ref{cmp} and Fig. \ref{g1}, we can observe that at time $t=100$ and $t=500$ probability using the proposed method is slightly higher compared to other techniques. Also for time $t=2500$ results derived using the proposed approach are slightly lower as compared to results derived from other techniques. Unlike traditional methods that rely on crisp data, our proposed technique accounts for uncertainty, providing a more accurate probability estimation of TE.

\begin{table}[H]
    \centering 
    \begin{tabular}{c|c|c|c} \hline
        t & Proposed Method & Petri-net & Bayesian Network \\ \hline
        100  & 0.05269844 & 0.04992134 & 0.04527448 \\
        500  & 0.56162025 & 0.55645041 & 0.52900833 \\
        1000 & 0.85253225 & 0.87518982 & 0.85004877 \\
        1500 & 0.94286331 & 0.96217232 & 0.94597227 \\
        2000 & 0.97599884 & 0.98690631 & 0.97782722 \\
        2500 & 0.98951921 & 0.99502308 & 0.99029561 \\
        3000 & 0.99532702 & 0.99800983 & 0.99563683 \\ \hline
    \end{tabular}
    \caption{Probability of TE using different approaches}
    \label{cmp}
\end{table}

\section{Conclusion}
While FTA is a highly effective and commonly used technique for dependability analysis, it does possess several limitations, notably its inability to capture sequential failure behavior. TFTA has been developed as an extension to overcome this limitation. TFTA introduces gates such as PAND and POR to address this issue. In this paper, we have extended the behavior of these gates in a fuzzy environment. We have developed an algorithm to compute TE probability using fuzzy failure rates of BE. We applied our extended formula to a case study of AFDS and compared our results with results derived from Petri net-based solutions and Bayesian network-based solutions. Importance measures enable analysts to assess the relative impact of individual system components on overall failure, without requiring precise failure probability estimates. The fuzzy approach facilitates meaningful conclusions about system failure behavior, bridging the gap created by uncertain data.

The current application of the method assumes exponentially distributed failure probabilities for components. Exploring alternative approaches to incorporate non-traditional failure distributions in quantitative assessments is a valuable direction for future research. In future, we aim to expand this research by investigating how the estimated system unreliability, obtained through the proposed approach, may be influenced by the various choices of membership functions.

\


\end{document}